\documentclass[10pt,a4paper]{amsart}
\usepackage{amsthm,amscd,times,amssymb,amsmath,diagrams}
\def\marg{}
\newtheoremstyle{SmallCaps}{}{}{\itshape}{}{\textsc\bgroup}{.\egroup}{
 }{}
\newcounter{theorem}
\theoremstyle{SmallCaps}
\newtheorem{dfn}[theorem]{Definition}
\newtheorem{cor}[theorem]{Corollary}
\newtheorem{lem}[theorem]{Lemma}

\newtheorem{exa}[theorem]{Example}
\newtheorem{thm}[theorem]{Theorem}

\newcommand{\cref}[1]{Corollary~\ref{#1}}
\newcommand{\lref}[1]{Lemma~\ref{#1}}

\newcommand{\eref}[1]{Example~\ref{#1}}
\newcommand{\tref}[1]{Theorem~\ref{#1}}
\def\please{\renewcommand\labelstyle{\scriptstyle}}
\def\R{\mathbb{R}}

\def\N{\mathbb{N}}

\def\One{\mathbb{I}}
\def\F{\mathbb{F}}
\def\D{\mathcal{D}}
\def\Lhi{\mathcal{L}}

\def\rhd{\succsim}
\def\ul{\underline}

\def\h{\mathcal{H}}
\def\htwo{\mathcal{H}^{\bullet\ast}}
\def\htwobar{\tilde{\mathcal{H}}^{\bullet\ast}}
\def\b{B_{+}}
\def\bbar{\tilde{B}_{+}}
\begin{document}

\def\oct{                            
\parbox{10mm}{\begin{picture}(10,15) 
\put(4.5,12){$\bullet$}              
\put(1,5){$\bullet$}
\put(8,5){$\circ$}
\put(5.5,13){\line(1,-2){3.3}}
\put(5.5,13){\line(-1,-2){3.3}}
\put(1,2){$\ul t$}
\end{picture}}}

\def\oline{                          
\parbox{3mm}{\begin{picture}(5,10)   
\put(1,8){$\bullet$}                 
\put(1,1){$\bullet$}
\put(2,9){\line(0,-1){7}}
\end{picture}}}

\def\olinephipsi{                    
\parbox{10mm}{\begin{picture}(10,10)  
\put(5,8){$$}                        
\put(5,1){$$}
\put(7.25,9){\line(0,-1){7}}
\put(2,1){$\phi(\bullet)$}
\put(1.5,8){$\psi(\bullet)$}
\end{picture}}}

\def\octline{                        
\parbox{3mm}{\begin{picture}(5,10)   
\put(1,8){$\ast   $}                 
\put(1,1){$\bullet$}
\put(2,9){\line(0,-1){7}}
\end{picture}}}

\def\ocbline{                        
\parbox{3mm}{\begin{picture}(5,10)   
\put(1,8){$\bullet$}                 
\put(1,1){$\ast$}
\put(2,9){\line(0,-1){7}}
\end{picture}}}

\def\octbline{                       
\parbox{3mm}{\begin{picture}(5,10)   
\put(1,8){$\ast   $}                 
\put(1,1){$\ast$}
\put(2,9){\line(0,-1){7}}
\end{picture}}}

\def\oc{                     
\parbox{10mm}{\begin{picture}(10,10) 
\put(4.5,8){$\bullet$}               
\put(1,1){$\bullet$}
\put(8,1){$\bullet$}
\put(5.5,9){\line(1,-2){3.5}}
\put(5.5,9){\line(-1,-2){3.5}}
\end{picture}}}

\def\ocn{                    
\parbox{10mm}{\begin{picture}(10,10) 
\put(4.5,8){$\bullet$}               
\put(1,1){$\circ$}
\put(8,1){$\circ$}
\put(5.5,9){\line(1,-2){3.3}}
\put(5.5,9){\line(-1,-2){3.3}}
\end{picture}}}

\def\ocnr{                       
\parbox{10mm}{\begin{picture}(10,10) 
\put(4.5,8){$\bullet$}               
\put(1,1){$\bullet$}
\put(8,1){$\circ$}
\put(5.5,9){\line(1,-2){3.3}}
\put(5.5,9){\line(-1,-2){3.3}}
\end{picture}}}

\def\othree{                     
\parbox{10mm}{\begin{picture}(10,10) 
\put(4.5,8){$\bullet$}               
\put(1,1){$\bullet$}
\put(4.5,1){$\bullet$}
\put(8,1){$\bullet$}
\put(5.5,9){\line(1,-2){3.5}}
\put(5.5,9){\line(-1,-2){3.5}}
\put(5.5,9){\line(0,-1){7}}
\end{picture}}}

\def\ofoura{                     
\parbox{13mm}{\begin{picture}(13,17) 
\put(8,15){$\bullet$}                
\put(4.5,8){$\bullet$}               
\put(8,8){$\bullet$}                 
\put(11.5,8){$\bullet$}
\put(1,1){$\bullet$}
\put(8,1){$\bullet$}
\put(9,16){\line(1,-2){3.5}}
\put(9,16){\line(-1,-2){3.5}}
\put(5.5,9){\line(1,-2){3.5}}
\put(5.5,9){\line(-1,-2){3.5}}
\put(9,16){\line(0,-1){7}}
\end{picture}}}

\def\occ{                     
\parbox{10mm}{\begin{picture}(10,17)  
\put(4.5,15){$\bullet$}               
\put(1,8){$\bullet$}                  
\put(8,8){$\bullet$}                  
\put(1,1){$\bullet$}
\put(5.5,16){\line(1,-2){3.5}}
\put(5.5,16){\line(-1,-2){3.5}}
\put(2,9){\line(0,-1){7}}
\end{picture}}}

\def\ofork{                   
\parbox{10mm}{\begin{picture}(10,17)  
\put(4.5,15){$\bullet$}               
\put(4.5,8){$\bullet$}                
\put(8,1){$\bullet$}                  
\put(1,1){$\bullet$}
\put(5.5,9){\line(0,1){7}}
\put(5.5,9){\line(-1,-2){3.5}}
\put(5.5,9){\line(1,-2){3.5}}
\end{picture}}}

\def\olongline{                   
\parbox{3mm}{\begin{picture}(3,17)    
\put(1,15){$\bullet$}                 
\put(1,8){$\bullet$}                  
\put(1,1){$\bullet$}                  
\put(2,2){\line(0,1){14}}
\end{picture}}}

\def\ocleft{                         
\parbox{10mm}{\begin{picture}(10,10) 
\put(4.5,8){$\bullet$}               
\put(1,1){$\ast$}
\put(8,1){$\bullet$}
\put(5.5,9){\line(1,-2){3.5}}
\put(5.5,9){\line(-1,-2){3.5}}
\end{picture}}}

\def\ocright{                        
\parbox{10mm}{\begin{picture}(10,10) 
\put(4.5,8){$\bullet$}               
\put(1,1){$\bullet$}
\put(8,1){$\ast$}
\put(5.5,9){\line(1,-2){3.5}}
\put(5.5,9){\line(-1,-2){3.5}}
\end{picture}}}

\def\ocleftright{                    
\parbox{10mm}{\begin{picture}(10,10) 
\put(4.5,8){$\bullet$}               
\put(1,1){$\ast$}
\put(8,1){$\ast$}
\put(5.5,9){\line(1,-2){3.5}}
\put(5.5,9){\line(-1,-2){3.5}}
\end{picture}}}

\def\octop{                          
\parbox{10mm}{\begin{picture}(10,10) 
\put(4.5,8){$\ast$}                  
\put(1,1){$\bullet$}
\put(8,1){$\bullet$}
\put(5.5,9){\line(1,-2){3.5}}
\put(5.5,9){\line(-1,-2){3.5}}
\end{picture}}}

\def\ocntop{                          
\parbox{10mm}{\begin{picture}(10,10) 
\put(4.5,8){$\ast$}                  
\put(1,1){$\circ$}
\put(8,1){$\circ$}
\put(5.5,9){\line(1,-2){3.3}}
\put(5.5,9){\line(-1,-2){3.3}}
\end{picture}}}

\def\oclefttop{                      
\parbox{10mm}{\begin{picture}(10,10) 
\put(4.5,8){$\ast$}                  
\put(1,1){$\ast$}
\put(8,1){$\bullet$}
\put(5.5,9){\line(1,-2){3.5}}
\put(5.5,9){\line(-1,-2){3.5}}
\end{picture}}}

\def\ocrighttop{                     
\parbox{10mm}{\begin{picture}(10,10) 
\put(4.5,8){$\ast$}                  
\put(1,1){$\bullet$}
\put(8,1){$\ast$}
\put(5.5,9){\line(1,-2){3.5}}
\put(5.5,9){\line(-1,-2){3.5}}
\end{picture}}}

\def\ocleftrighttop{                 
\parbox{10mm}{\begin{picture}(10,10) 
\put(4.5,8){$\ast$}                  
\put(1,1){$\ast$}
\put(8,1){$\ast$}
\put(5.5,9){\line(1,-2){3.5}}
\put(5.5,9){\line(-1,-2){3.5}}
\end{picture}}}

\def\olone{                          
\parbox{3mm}{\begin{picture}(5,10)   
\put(1,8){$\bullet_1$}               
\put(1,1){$\bullet$}
\put(2,9){\line(0,-1){7}}
\end{picture}}}

\def\otwo{                         
\parbox{3mm}{\begin{picture}(5,10) 
\put(1,8){$\bullet_2$}             
\put(1,1){$\bullet$}
\put(2,9){\line(0,-1){7}}
\end{picture}}}

\def\ocnp{                             
\parbox{16mm}{\begin{picture}(16,17)   
\put(4.5,8){$\bullet$}                 
\put(1,1){$\circ$}                     
\put(8,1){$\circ$}                     
\put(5.5,9){\line(1,-2){3.3}}
\put(5.5,9){\line(-1,-2){3.3}}
\put(8,15){$\bullet$}
\put(15,1){$\circ$}
\put(9,16){\line(-1,-2){3.5}}
\put(9,16){\line(1,-2){6.7}}
\end{picture}}}

\def\ocnpo{                            
\parbox{16mm}{\begin{picture}(16,17)   
\put(4.5,8){$\bullet$}                 
\put(1,1){$\bullet$}                   
\put(8,1){$\bullet$}                   
\put(5.5,9){\line(1,-2){3.3}}
\put(5.5,9){\line(-1,-2){3.3}}
\put(8,15){$\bullet$}
\put(15,1){$\bullet$}
\put(9,16){\line(-1,-2){3.5}}
\put(9,16){\line(1,-2){6.7}}
\end{picture}}}

\def\ocnplr{                           
\parbox{16mm}{\begin{picture}(16,17)   
\put(4.5,8){$\bullet_{12}$}            
\put(1,1){$\bullet_1$}                 
\put(8,1){$\bullet_2$}                 
\put(5.5,9){\line(1,-2){3.5}}
\put(5.5,9){\line(-1,-2){3.5}}
\put(8,15){$\bullet_{123}$}
\put(15,1){$\bullet_3$}
\put(9,16){\line(-1,-2){3.5}}
\put(9,16){\line(1,-2){7}}
\end{picture}}}

\def\ocnprl{                           
\parbox{16mm}{\begin{picture}(16,17)   
\put(11.5,8){$\bullet_{23}$}           
\put(1,1){$\bullet_1$}                 
\put(8,1){$\bullet_2$}                 
\put(12.5,9){\line(1,-2){3.5}}
\put(12.5,9){\line(-1,-2){3.5}}
\put(8,15){$\bullet_{123}$}
\put(15,1){$\bullet_3$}
\put(9,16){\line(1,-2){3.5}}
\put(9,16){\line(-1,-2){7}}
\end{picture}}}

\def\ocnpt{                            
\parbox{16mm}{\begin{picture}(16,17)   
\put(4.5,8){$\bullet$}                 
\put(1,1){$\circ$}                     
\put(8,1){$\circ$}                     
\put(5.5,9){\line(1,-2){3.3}}
\put(5.5,9){\line(-1,-2){3.3}}
\put(8,15){$\ast$}
\put(15,1){$\circ$}
\put(9,16){\line(-1,-2){3.5}}
\put(9,16){\line(1,-2){6.7}}
\end{picture}}}

\def\ocnptb{                           
\parbox{16mm}{\begin{picture}(16,17)   
\put(4.5,8){$\ast   $}                 
\put(1,1){$\circ$}                     
\put(8,1){$\circ$}                     
\put(5.5,9){\line(1,-2){3.3}}
\put(5.5,9){\line(-1,-2){3.3}}
\put(8,15){$\ast$}
\put(15,1){$\circ$}
\put(9,16){\line(-1,-2){3.5}}
\put(9,16){\line(1,-2){6.7}}
\end{picture}}}

\def\ocnpb{                            
\parbox{16mm}{\begin{picture}(16,17)   
\put(4.5,8){$\ast   $}                 
\put(1,1){$\circ$}                     
\put(8,1){$\circ$}                     
\put(5.5,9){\line(1,-2){3.3}}
\put(5.5,9){\line(-1,-2){3.3}}
\put(8,15){$\bullet$}
\put(15,1){$\circ$}
\put(9,16){\line(-1,-2){3.5}}
\put(9,16){\line(1,-2){6.7}}
\end{picture}}}
\def\osix{                              
\parbox{37mm}{\begin{picture}(37,24)    
\put(4.5,8){$\bullet$}                  
\put(1,1){$\bullet$}                    
\put(8,1){$\bullet$}                    
\put(5.5,9){\line(1,-2){3.5}}           
\put(5.5,9){\line(-1,-2){3.5}}
\put(18.5,8){$\bullet$}
\put(15,1){$\bullet$}
\put(22,1){$\bullet$}
\put(19.5,9){\line(1,-2){3.5}}
\put(19.5,9){\line(-1,-2){3.5}}
\put(32.5,8){$\bullet$}
\put(29,1){$\bullet$}
\put(36,1){$\bullet$}
\put(33.5,9){\line(1,-2){3.5}}
\put(33.5,9){\line(-1,-2){3.5}}
\put(18.5,22){$\bullet$}
\put(11.5,15){$\bullet$}
\put(19.5,9){\line(-1,1){7}}
\put(12.5,16){\line(-1,-1){7}}
\put(19.5,23){\line(-1,-1){7}}
\put(19.5,23){\line(1,-1){14}}
\end{picture}}}

\def\osixn{                 
\parbox{37mm}{\begin{picture}(37,24)    
\put(4.5,8){$\bullet$}                  
\put(1,1){$\circ$}                      
\put(8,1){$\circ$}                      
\put(5.5,9){\line(1,-2){3.3}}           
\put(5.5,9){\line(-1,-2){3.3}}
\put(18.5,8){$\bullet$}
\put(15,1){$\circ$}
\put(22,1){$\circ$}
\put(19.5,9){\line(1,-2){3.3}}
\put(19.5,9){\line(-1,-2){3.3}}
\put(32.5,8){$\bullet$}
\put(29,1){$\circ$}
\put(36,1){$\circ$}
\put(33.5,9){\line(1,-2){3.3}}
\put(33.5,9){\line(-1,-2){3.3}}
\put(18.5,22){$\bullet$}
\put(19.5,23){\line(0,-1){14}}
\put(12.5,16){\line(-1,-1){7}}
\put(19.5,23){\line(-1,-1){7}}
\put(19.5,23){\line(1,-1){14}}
\end{picture}}}

\def\osixend{              
\parbox{37mm}{\begin{picture}(37,24)    
\put(4.5,8){$\bullet$}                  
\put(1,1){$\bullet$}                    
\put(8,1){$\bullet$}                    
\put(5.5,9){\line(1,-2){3.5}}           
\put(5.5,9){\line(-1,-2){3.5}}
\put(18.5,8){$\bullet$}
\put(15,1){$\bullet$}
\put(22,1){$\bullet$}
\put(19.5,9){\line(1,-2){3.5}}
\put(19.5,9){\line(-1,-2){3.5}}
\put(32.5,8){$\bullet$}
\put(29,1){$\bullet$}
\put(36,1){$\bullet$}
\put(33.5,9){\line(1,-2){3.5}}
\put(33.5,9){\line(-1,-2){3.5}}
\put(18.5,22){$\bullet$}
\put(19.5,23){\line(0,-1){14}}
\put(12.5,16){\line(-1,-1){7}}
\put(19.5,23){\line(-1,-1){7}}
\put(19.5,23){\line(1,-1){14}}
\end{picture}}}

\def\osixnum{               
\parbox{80mm}{\begin{picture}(80,24)    
\put(4.5,8){$\bullet_{12}$}             
\put(1,1){$\bullet_1$}                  
\put(8,1){$\bullet_2$}                  
\put(5.5,9){\line(1,-2){3.5}}           
\put(5.5,9){\line(-1,-2){3.5}}
\put(18.5,8){$\bullet_{34}$}
\put(15,1){$\bullet_3$}
\put(22,1){$\bullet_4$}
\put(19.5,9){\line(1,-2){3.5}}
\put(19.5,9){\line(-1,-2){3.5}}
\put(32.5,8){$\bullet_{56}$}
\put(29,1){$\bullet_5$}
\put(36,1){$\bullet_6$}
\put(33.5,9){\line(1,-2){3.5}}
\put(33.5,9){\line(-1,-2){3.5}}
\put(11.5,15){$\bullet_{1234}$}
\put(18.5,22){$\bullet_{123456}$}
\put(12.5,16){\line(1,-1){7}}
\put(12.5,16){\line(-1,-1){7}}
\put(19.5,23){\line(-1,-1){7}}
\put(19.5,23){\line(1,-1){14}}
\put(50,1){- level 0}
\put(50,8){- level 1}
\put(50,15){- level 2}
\put(50,22){- level 3}
\end{picture}}}

\def\ocfive{                         
\parbox{37mm}{\begin{picture}(37,24) 
\put(4.5,8){$\ast$}                  
\put(1,1){$\circ$}                   
\put(8,1){$\circ$}                   
\put(5.5,9){\line(1,-2){3.3}}        
\put(5.5,9){\line(-1,-2){3.3}}       
\put(18.5,8){$\ast$}
\put(15,1){$\circ$}
\put(22,1){$\circ$}
\put(19.5,9){\line(1,-2){3.3}}
\put(19.5,9){\line(-1,-2){3.3}}
\put(32.5,8){$\bullet$}
\put(29,1){$\circ$}
\put(36,1){$\circ$}
\put(33.5,9){\line(1,-2){3.3}}
\put(33.5,9){\line(-1,-2){3.3}}
\put(11.5,15){$\bullet$}
\put(18.5,22){$\ast$}
\put(12.5,16){\line(1,-1){7}}
\put(12.5,16){\line(-1,-1){7}}
\put(19.5,23){\line(-1,-1){7}}
\put(19.5,23){\line(1,-1){14}}
\end{picture}}}

\def\ocful{
\parbox{16mm}{\begin{picture}(16,15)
\put(8,12){$\bullet$}
\put(1,5){$\bullet$}
\put(15,5){$\bullet$}
\put(9,13){\line(1,-1){7}}
\put(9,13){\line(-1,-1){7}}
\put(9,13){\line(1,-2){3.5}}
\put(9,13){\line(0,-1){7}}
\put(9,13){\line(-1,-2){3.5}}
\put(1,2){${\ul t}_1$}
\put(7,2){$\ldots$}
\put(15,2){${\ul t}_n$}
\end{picture}}}

\title{The Hopf algebra of rooted trees in Epstein-Glaser renormalization}

\author{Christoph Bergbauer and Dirk Kreimer}

\address{II. Mathematisches Institut\\Fachbereich Mathematik und Informatik\\Freie Universit\"at Berlin\\Arnimallee 3\\D-14195 Berlin and \newline Institut des Hautes \'Etudes Scientifiques\\35 route de Chartres\\F-91440 Bures-sur-Yvette}
\email{bergbau@math.fu-berlin.de}

\address{CNRS at Institut des Hautes \'Etudes Scientifiques\\35 route de Chartres\\F-91440 Bures-sur-Yvette and
Center for Mathematical Physics, Boston University} \email{kreimer@ihes.fr}

\date{}

\begin{abstract}
We show how the Hopf algebra of rooted trees encodes the combinatorics of Epstein-Glaser renormalization and coordinate space renormalization in general. In particular, we prove that the Epstein-Glaser time-ordered products can be obtained from the Hopf algebra by suitable Feynman rules, mapping trees to operator-valued distributions. Twisting the antipode with a renormalization map formally solves the Epstein-Glaser recursion and provides local counterterms due to the Hochschild 1-closedness of the grafting operator $B_+$.
\end{abstract}

\maketitle

\section*{Introduction}
The Epstein-Glaser framework \cite{EG,Scharf} and its modern variants \cite{PS,BF,Pinter} provide a mathematically rigorous approach to perturbation theory and
renormalization in coordinate space. Let $M=\R^{1,3}$ denote the Minkowski space. Epstein and Glaser constructed, for a scalar $\phi^k$ field theory say, a sequence of operator-valued distributions $T_n$ on $M^n$
respectively, which replace the ill-defined time-ordered products in the standard approach to perturbation theory. The result
is a perturbation theory which is a priori finite in each order - no removal of short-distance singularities is needed since all expressions are well-defined
from the very beginning.  The appropriate notion of renormalization in the Epstein-Glaser framework is \emph{extension
  of distributions onto diagonals}. Indeed, the objects of interest $T_n$
are a priori determined outside the diagonals by causality. Finite renormalizations correspond to different ways of extending distributions onto diagonals. Moreover, in this approach the S-matrix is \emph{local} by construction.\\\\
On the other hand, the combinatorics of momentum space renormalization have been most efficiently described \cite{CK2,Kreimer3} in terms of the Hopf algebra and associated Lie algebra of Feynman graphs.
Renormalization and in particular the Bogoliubov recursion boil down to twisting the antipode $S$ of that Hopf algebra by renormalization maps into some target ring of Laurent or formal power series. This is possible due to a coproduct which disentangles 1PI graphs into divergent 1PI subgraphs. There is a universal object behind all Hopf algebras of this kind: the Hopf algebra of rooted trees \cite{Kreimer,CK} which encodes nested subdivergences in terms of a tree and their recursive removal in terms of its coproduct and the resulting antipode. We will show how the Hopf algebra of rooted trees works in the realm of Epstein-Glaser renormalization in almost complete analogy to other renormalization programs like BPHZ. In fact it is even easier to understand its role in Epstein-Glaser renormalization since no regularization is required and overlapping divergences do not exist in the coordinate space language. \\\\
This paper is organized as follows: In the first section we give a short review of the Epstein-Glaser construction of time-ordered products, emphasizing the point of view of diagonals \cite{BF}. The second section recalls the powerful notion of a Hochschild 1-cocycle on a connected graded bialgebra, giving rise to two equivalent presentations of the Hopf algebra of rooted trees. A new convolution-like product is introduced which in cooperation with the antipode allows to recursively generate all terms needed for an Epstein-Glaser time-ordered product, as will be proved using explicit renormalized ``Feynman rules'' in the final theorem which we already state in a short version: \\\\
\sc Theorem  \rm (Main result) \it There is a map $\Phi: \mathcal{H}^{\bullet\ast} \rightarrow FV$ such that the $n$-th Epstein-Glaser time-ordered product $T_n$ is given by
\begin{equation*}
T_n = \sum_{t\in \mathcal{T}_n} \Phi(S_R\odot id)(t)
\end{equation*}
\rm where $\mathcal{H}^{\bullet\ast}$ is a Hopf-algebra of rooted trees, $FV$ something like the tensor algebra of distributions on $M$, $\mathcal{T}_n$ the set of all binary trees with $n$ leaves, $S_R$ the twisted antipode of $\mathcal{H}^{\bullet\ast}$ and $\odot$ a modified ``convolution product'' in $\htwo.$

\section{Some background on Epstein-Glaser renormalization}
For simplicity we restrict ourselves to a massive neutral scalar field theory 
with interaction
Lagrangian
\begin{equation}
\label{Lagrangian}
\Lhi_I=\frac{\lambda}{k!}\phi^k,
\end{equation}
on the flat Minkowski space-time $M:=\R^{1,3}.$ Generalizations to
Quantum Electrodynamics and globally hyperbolic space-times have been
worked out in \cite{Scharf} and \cite{BF}, respectively, which though does not
affect the combinatorics we are primarily interested in.

\subsection{Motivation}
As a starting point for the Epstein-Glaser construction of time-ordered products \cite{EG} we
consider the symbolic Dyson series for the S-matrix
\begin{equation}\label{Dyson}
S =  T e ^{i\int\Lhi_I (x) dx}
\end{equation}
which is formally derived from the Schwinger differential equation of motion by transforming it
into an iterated integral equation and applying the time-ordering
operator $T$ to each summand
\begin{equation*}
\frac{i^n}{n!}\int_{M^n} T(\Lhi_I(x_1)\ldots \Lhi_I(x_n))dx_1\ldots dx_n
\end{equation*}
which has the benefit that we are integrating now over $M^n$ rather than over an
$n\mbox{-simplex }\times\R^{3n}.$\\\\
Let $A,B$ be operator-valued functions on $M.$
The time-ordering operator $T$ is usually defined by
\begin{equation}
\label{oldt}
T(A(x_1) B(x_2)) :=\Theta(x_{1}^0-x_{2}^0)A(x_1)B(x_2)+\Theta(x_{2}^0-x_{1}^0)B(x_2)A(x_1)
\end{equation}
where $\Theta$ denotes the Heaviside characteristic function of $\R_{\ge 0}$. Analogously one defines $T$ on more than two factors. \\\\
Now $S$ and $\Lhi_I$ are obviously supposed to be
operator-valued \emph{distributions}, for which (\ref{oldt}) does not
make sense since distributions can not just be multiplied by
noncontinuous functions like $\Theta.$ It does make sense though outside the
thick diagonal $D_n=\{x\in M^n: x_i=x_j$ for some $i\neq j\}$ where products of
$\Theta(x_i^0-x_j^0)$ are continuous. \\\\
In fact the mathematical origin for the appearance of short-distance singularities
in perturbation theory is the ill-defined notion of time-ordering reviewed above. Epstein
and Glaser proposed a way to construct well-defined time ordered products $T_n,$
one for each power $n$ of the coupling constant, that satisfy a set
of suitable conditions explained below, the most prominent being
that of \emph{locality} or \emph{micro-causality}. The power series $S$ constructed by (\ref{Dyson}) using the Epstein-Glaser time-ordered product $T$ is a priori finite in every order, and
renormalization corresponds then to stepwise \emph{extension of
  distributions} from $M^n-D_n$ to $M^n.$ In general, distributions can not be extended uniquely onto diagonals. The resulting degrees of freedom are in one-to-one correspondence with the degrees of freedom (finite renormalizations) in momentum space renormalization programs like BPHZ and dimensional regularization.  \\\\
The notion of \emph{locality}, crucial to the following construction
of time-ordered products, can be motivated as follows: Suppose
$x=(x_1,\ldots,x_n)\in M^n,$ $\emptyset\subsetneq I\subsetneq N:=\{1,\ldots, n\}$ and for each $i\in I,$ the point $x_i$ is not in
the past causal shadow of any of the $x_j$ for $j\in N-I.$ We denote this situation
$x_i\rhd x_j$ $\forall i\in I,j\in N-I.$ Then our time ordered product $T_{n}$ is supposed to satisfy (in the sense of operator-valued distributions)
\begin{equation}
\label{naivecausality}
T_{n}(x_1,\ldots,x_n)=T_{|I|}(x_i)_{i\in I}T_{|N-I|}(x_j)_{j\in N-I}
\end{equation}
because we think of the $x_i$ to happen after (or at least not before) the $x_j.$
If both $x_i\rhd x_j$ and $x_j\rhd x_i,$ $\forall i,j,$ so if all pairs $(x_i,x_j)$ are
spacelike,
we have $[T_{|I|}(x_i)_{i\in I},T_{|N-I|}(x_j)_{j\in N-I}]=0.$

\subsection{Construction of time-ordered products}
In this subsection we give a short review of the mathematical core of Epstein-Glaser
renormalization in its modern variant \cite{PS,BF,Pinter} which
emphasizes the point of view of nested diagonals. For
the proofs, the reader is referred to \cite{BF}. \\\\
The Minkowski metric on $M$ provides a relation $\rhd$ on $M$ as follows: $x\rhd y$ iff $x$ is not in the past causal shadow
of $y,$ that is $x\notin y+\overline{V}^-$ where $\overline{V}^-:=\{z\in M: (z)^2\le 0,\,
z^0 \le 0\}$ is the closed past lightcone. \\\\
Now, for $n\in\N$ let $N:=\{1,\ldots,n\}$ and $\emptyset\subsetneq I\subsetneq N.$ The set
\begin{equation*}
C_I := \{(x_1,\ldots,x_n)\in M^n:\, x_i\rhd x_j\,\forall i\in I,\,
j\in N-I\}
\end{equation*}
is obviously a translation invariant open subset of $M^n.$
\begin{lem}[Geometric lemma]
\label{Geometric lemma}
\begin{equation*}
\bigcup_{\emptyset\subsetneq I\subsetneq N} C_I = M^n-\Delta_n
\end{equation*} where
$\Delta_n =\{x\in M^n:\, x_1=\ldots=x_n\}$ is the ``thin'' diagonal in $M^n.$
\end{lem}
The proof is an easy induction on $n.$
The geometric lemma tells us that the causality condition (\ref{naivecausality})
determines the time-ordered product $T_n$ everywhere
outside the thin diagonal $\Delta_n,$ once the $T_k$ for $k< n$ are
known on whole $M^k,$ respectively. It is important to understand that the geometric lemma does not really constitute a specific feature of the Minkowski space. Indeed, the lemma holds if one replaces $\rhd$ by any relation such that $x\rhd y$ or $y\rhd x$ whenever $y\neq x,$ and such that $\rhd$ is ``weakly transitive'' 
in the sense that $x\rhd y$ and $\neg (z\rhd y)$ imply $x\rhd z.$
\begin{dfn} A \emph{causal partition of unity} $\{p_{I,N-I}\}_{\emptyset\subsetneq I\subsetneq N}$ is a smooth partition
  of unity subordinate to the cover $\{C_{I}\}_{\emptyset\subsetneq I\subsetneq N}$ of $M^n-\Delta_n.$
\end{dfn}
For simplicity, we will sometimes drop the curly brackets in the subscript, for example $p_{1,2}$ denotes $p_{\{1\},\{2\}}.$\\\\
Let $\D(M)=C_0^\infty(M)$ denote the space of test functions on $M$ with the usual topology. Let $H$ denote the Hilbert space of the free field theory and $D$ a suitable dense subspace.
In principle an Epstein-Glaser time-ordered
product is a collection $(T^r_n)_{n\in\N}$ ($r=(r_1,\ldots,r_n)$ an $n$-multiindex) of operator-valued distributions
$T^r_n: \D(M^n)\rightarrow End(D),$ such that $T^{(r_1,\ldots,r_n)}_n$ replaces the time ordering of the $n$ Wick monomials
$:\phi^{r_1}:,\ldots,:\phi^{r_n}:.$
\begin{dfn}
\label{time-ordered product}
A collection $(T^r_n)$ of operator-valued distributions
$T^r_n: \mathcal{D}(M^n)\rightarrow End(D)$
is called an (Epstein-Glaser) time-ordered product
  if
\begin{enumerate}
\item[(i)] $T^k_1(f)=\,:\phi^k:(f)$ where $:\phi^k:(f)$ denotes the Wick monomial $:\phi^k:$ smeared with the test function $f,$
\item[(ii)] $T$ is symmetric
\begin{equation*}
T^r_n(f_1\otimes\ldots\otimes f_n)=T^r_n(f_{\pi(1)}\otimes\ldots\otimes f_{\pi(n)})
\end{equation*}
when $\pi$ is a permutation of $N:=\{1,\ldots,n\}.$ This allows for the
notation
\begin{equation*}
T(N)=T^r_n(f_1\otimes\ldots\otimes f_n)
\end{equation*}
when the $f_i$ and $r_i$ are clear from the context,
\item[(iii)] $T$ splits causally: Let $\emptyset \subsetneq I\subsetneq N.$ Then
\begin{equation}
\label{causality}
T(N)=T(I)T(N-I)\\
\end{equation}
for all test functions \emph{with support in $C_I\subset M^n,$}
\item[(iv)] $T$ is translation covariant
\begin{equation*}
  U(a,1)T(f_1,\ldots,f_n)U(a,1)^{-1}=T(\tau_a f_1,\ldots,\tau_a f_n)
\end{equation*}
where $U(\cdot,1)\ldots U(\cdot,1)^{-1}$ is the representation of the translation part of the Poincar\'e group in $D,$ and $\tau_af(x)=f(x-a)$ denotes translation by $a.$
\item[(v)] The Wick expansion relates time-ordered products corresponding to different Wick-powers
\begin{eqnarray}\label{wick}
T_n^{(r_1,\ldots,r_n)}(f_1\otimes\ldots\otimes f_n)&=&\sum_{i_1,\ldots,i_n}\left<\Omega,T_n^{(r_1-i_1,\ldots,r_n-i_n)}(f_1\otimes\ldots\otimes f_n)\Omega\right>\nonumber\\ &&\times {r_1 \choose i_1}\ldots {r_n\choose i_n}:\phi^{i_1}\ldots\phi^{i_n}:(f_1\otimes\ldots\otimes f_n)
\end{eqnarray}
with $\Omega$ the vacuum state in $D\subset H.$
\end{enumerate}
\end{dfn}
Note that by the so called Theorem 0 in \cite{EG} the summands in the right hand side of (\ref{wick}) as products of translation invariant \emph{numerical} distributions and Wick monomials are well-defined operator-valued distributions. Once a time-ordered product $T=(T^r_n)$ is given, the S-matrix for the $\phi^k$-theory is obtained as the formal power series
\begin{equation}\label{SfromT}
S(f)=\sum_{n=0}^\infty \frac{i^n}{n!(k!)^n}T^{(k,\ldots,k)}_n(f^{\otimes n}),
\end{equation}
\marg
possibly taking the adiabatic limit $f\rightarrow \lambda$ later on, which is a highly nontrivial task we shall not be concerned about in the present work. The S-matrix (\ref{SfromT}) and the relative S-matrices constructed from $T$ are local. If one imposes additional normalization conditions (Lorentz covariance, Hermiticity etc., see \cite{BF}) on $T,$ the S-matrix becomes Lorentz covariant and unitary, etc. Moreover, the interacting field constructed from the relative S-matrices are Lorentz covariant, Hermitean and satisfy the interacting field equation.   
\begin{thm}\label{existence}
Time-ordered products exist.
\end{thm}
A constructive proof is given in \cite{BF} and of course, but in a somewhat different notation, in the original paper \cite{EG}. The idea is as follows: Provided all $(T_m)$ for $m<n$ are constructed, the Geometric \lref{Geometric lemma} ensures that $T_n$ is determined on $M^n-\Delta_n$ by causality (iii). We define
\begin{equation*}
T_I=T(I)T(N-I)\mbox{ as a distribution on }C_I.
\end{equation*} One easily shows that
\begin{equation*}
T(I)T(N-I)=T(J)T(N-J)
\end{equation*}
on the intersection $C_I\cap C_J.$ Therefore, we can patch the $T_I$ together using a causal partition of unity $\{p_{I,N-I}\}$
\begin{equation}\label{top1}
^0T(N):=\sum_{\emptyset\subsetneq I\subsetneq N}p_{I,N-I}T(I)T(N-I)
\end{equation}
which is a well-defined distribution on $M^n-\Delta_n.$ As usual, $^0T(N)$ is independent on the choice of the partition of unity. It remains to extend it to a distribution on $M^n.$ Using the Wick expansion (v) and translation invariance, this amounts to an extension problem of numerical distributions $^0t_n$ from $M^{n-1}-\{0\}$ to $M^{n-1}.$ Having quantified the behavior of a numerical distribution at the origin by the \emph{Steinmann scaling degree} (see \cite{BF} for details), a generalization of the degree of homogeneity, one can show that there is a unique extension $t_n$ of $^0t_n$ to $M^{n-1}$ preserving the scaling degree, provided the scaling degree $sd(^0t_n)$ of $^0t_n$ is smaller than the dimension $4(n-1).$ Otherwise, if it is larger or equal but still finite, there is a finite dimensional space of extensions obtained as follows:
Let $f\in\D(M^{n-1}).$
The distribution
\begin{equation}
t_n: f\mapsto \left.^0t_n\right.\left(f-\sum_{\alpha}\omega_\alpha \partial^\alpha f(0)\right)
\end{equation}
where the sum goes over all $4(n-1)$-multiindices $\alpha$ such that $|\alpha|\le sd(^0t_n)-4(n-1)$ and the $\omega_\alpha\in\D(M^{n-1})$ such that $\partial^\beta\omega_\alpha(0)=\delta_{\alpha,\beta},$ has then scaling degree $sd(^0t_n)<4(n-1)$ and is hence uniquely extendible (preserving the scaling degree). There is an ambiguity due to the $\omega_\alpha$ however, and it is exactly this ambiguity which corresponds to the freedom of finite renormalizations. We call the linear operator $id-w$ on test functions
\begin{equation*}
id-w: f\mapsto f-\sum_{\alpha}\omega_\alpha \partial^\alpha f(0)
\end{equation*}
\emph{Taylor subtraction operator} and, motivated by the fact that
\begin{equation*}
t_n = (id-w^\ast) ^0t_n
\end{equation*}
holds on the level of numerical distributions, we write \emph{by abuse of notation} the extension of $^0T(N)$ to the diagonal by
\begin{equation}\label{top2}
T(N) = (id-W^{\ast}_{1\ldots n})^0T(N)
\end{equation}
although there is no linear operator $W^\ast$ on the space of operator valued distributions doing this duty. Our abuse of notation is justified though because we are only concerned with the combinatorics with respect to $n$ in the following, and the Wick expansion leaves $n$ obviously unchanged. So we understand $W^\ast$ as the symbolic ``operator'' which unpacks the operator valued distributions into Wick monomials and numerical distributions, Taylor subtracts the test function for those numerical distributions and produces then a ``counterterm'' such that $(id-W^\ast)$ maps a distribution on $M^n-\Delta_n$ to an extension on $M^n$ while the possible ambiguity (depending on the scaling degrees) is fixed by a choice of the $\omega_\alpha$.  The subscript in $W^{\ast}_{1\ldots n}$ indicating to which coordinates it applies will be useful later on. \\\\
This constructive proof of \tref{existence} actually proves more than the theorem demands: that in each extension step the scaling degree does not increase. If we make this an additional condition on time-ordered products, we can state
\begin{cor}\label{uniqueness}
All time-ordered products are uniquely (up to the $\omega_\alpha$, more precisely up to the finite set of constants $^0t_n(\omega_\alpha)$ in every order $n$) characterized by equations (\ref{top1}) and (\ref{top2}).
\end{cor}
\marg
Feynman graphs enter the game when one applies Wick's theorem. It might be instructive to have a look at the examples in \cite{Pinter}.
We also note that the usual notions of renormalizable theories, critical dimension etc.~can be traced back to the behavior of the scaling degrees as $n$ and the space-time dimension vary. In particular, the scaling degree coincides with the usual power-counting techniques in momentum space.

\section{The Hopf algebra of rooted trees in Epstein-Glaser
  renormalization}
The combinatorics of renormalization in coordinate space can be most
easily described in terms of rooted trees. Given some space-time points,
\begin{equation*}
\bullet\quad\bullet\quad\bullet\quad\bullet\quad\bullet\quad\bullet
\end{equation*}
we consider them as leaves of a tree (to be constructed). 
Whenever some of these points come together on a diagonal in $M^n,$ we connect the corresponding vertices to a new vertex such that subdivergences (subdiagonals) correspond to subtrees, for example
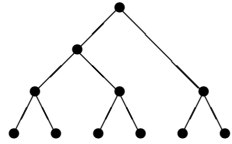
\begin{equation*}
\osix
\end{equation*}
So a tree represents the (partially ordered) nested or disjoint subdiagonals which are relevant to renormalization. It is now possible to construct a suitable coproduct on the free algebra generated by these trees such that the Bogoliubov recursion is essentially solved by the antipode of the resulting Hopf algebra on trees, as will be made precise in subsection \ref{normalhopf}. This remarkable property and the fact that local counterterms result \cite{Kreimer2} are the consequence of the fact that a certain operator on the Hopf algebra is a Hochschild 1-cocycle.
\subsection{Hochschild cohomology of bialgebras} \label{Hochschild}
All algebras are supposed to be over some field $k$ of characteristic zero, associative and unital, analogously for coalgebras. The unit (and by abuse of notation also the unit map) will be denoted by $\One, $ the counit map by $\epsilon.$ \emph{All algebra homomorphisms are supposed to be unital.} A bialgebra $\left(A=\bigoplus_{i=0}^\infty A_i,m,\One,\Delta,\epsilon\right)$ is called \emph{graded connected} if $A_iA_j\subset A_{i+j}$ and $\Delta(A_i)\subset\bigoplus_{j+k=i}A_j\otimes A_k,$ and if $\Delta(\One)=\One\otimes\One$ and $A_0=k\One,$ $\epsilon(\One)=\One$ and $\epsilon=0$ on $\bigoplus_{i=1}^\infty A_i.$ We call $\ker\epsilon$ the augmentation ideal of $A$ and denote $P$ the projection $A\rightarrow A$ onto the augmentation ideal, $P=id-\One\epsilon.$ \\\\
Let $(A,m,\One,\Delta,\epsilon)$ be a bialgebra. We think of linear maps $L:
A\rightarrow A^{\otimes n}$ as $n$-cochains and define a coboundary map $b$ by
\begin{equation}
\label{Hscb}
bL := (id\otimes L)\circ \Delta+\sum_{i=1}^n(-1)^i\Delta_i\circ
L+(-1)^{n+1}L\otimes \One
\end{equation}
where $\Delta_i$ denotes the coproduct applied to the $i$-th factor in $A^{\otimes n}$. It is easy
to see (using essentially the coassociativity of $\Delta$) that
$b^2=0,$ which gives rise to a cohomology theory called Hochschild cohomology.\\\\
It is also easy to see that, for $A$ finite dimensional say, the
cohomology theory (\ref{Hscb}) is the dual of the usual Hochschild
homology of the dual algebra $A^\ast.$\\\\
In case $n=1,$ (\ref{Hscb}) reduces to, for $L: A\rightarrow A,$
\begin{equation}
\label{Hscb1}
bL = (id\otimes L)\circ\Delta-\Delta\circ L+L\otimes\One.
\end{equation}
It is known \cite{CK} that the category of objects $(A,C)$
consisting of a commutative bialgebra $A$ and a Hochschild 1-cocycle $C$ on $A$
with morphisms bialgebra morphisms commuting with the cocycles has an
initial object $(\mathcal{H},B_+)$, with $\mathcal{H}$ the Hopf algebra of (non-planar) rooted
trees and the operator $B_+$ which grafts a product of rooted trees
together to a new root as described in the next subsection.\\
While the higher $(n>1)$ Hochschild cohomology of $\mathcal{H}$ vanishes
\cite{Foissy}, the closedness of $B_+$ will turn out to be crucial
for what follows.\\\\
The next lemma will provide a convenient way to construct Hopf algebras out of free or free commutative algebras by choosing linear endomorphisms $C_i$ and demanding that the $C_i$ be Hochschild 1-cocycles.  
\begin{lem} \label{bplusyieldshopf}
Let $A=\bigoplus_{n=0}^\infty A_n$ be a free or free commutative graded algebra (generated by a graded vector space) such that $A_0=k\One,$ and let $(C_i)_{i\in I}$ be a collection of injective linear endomorphisms of $A$ such that $C_i(A)\cap C_j(A)=\{0\}$ for $i\neq j$ and such that each free generator $y$ in degree $n$ is the image under some $C_i$ of an $x\in A_{n-1}$ for $n\ge 1.$ Then there is a unique connected graded bialgebra structure $(\Delta,\epsilon)$ on $A$ such that the $C_i$ are Hochschild closed with respect to $\Delta.$ In particular, $A$ is a Hopf algebra (with this property) in a unique way.
\end{lem}
\emph{Proof. } We will construct $\Delta$ by induction on $n.$ The Hochschild closedness of the $C_i$ demands that

\begin{equation}
\label{hoch}
\Delta\circ C_i = (id\otimes C_i)\circ\Delta +C_i\otimes\One.
\end{equation}
$\Delta(\One)=\One\otimes\One$ by convention, so $\Delta$ is known on $A_0.$ Now let $y$ be a free generator in $A_{n+1}.$ By assumption there is a unique $x\in A_n$ such that $y=C_i x$ for exactly one $i\in I.$ Assume $\Delta$ is known on $x$, then by (\ref{hoch}) it is also known on $y.$ Hence we can uniquely extend $\Delta$ to an algebra homomorphism on $A_{n+1}.$ By induction, this uniquely defines $\Delta$ as an algebra morphism on $A.$ From (\ref{hoch}) it also follows inductively that $\Delta$ respects the grading in all orders:
\begin{equation*}
\Delta(A_n)\subset\bigoplus_{k=0}^n A_k\otimes A_{n-k}.
\end{equation*} \\
For the coassociativity $(\Delta\otimes id)\Delta=(id\otimes\Delta)\Delta$ we note that
\begin{eqnarray*}
(\Delta\otimes id)\Delta C_i&=&(\Delta\otimes id)((id\otimes C_i)\Delta+C_i\otimes\One)\\
 &=&(\Delta\otimes C_i)\Delta+\Delta C_i\otimes\One\\
 &=&(\Delta\otimes C_i)\Delta+(id\otimes C_i\otimes id)(\Delta\otimes\One)+C_i\otimes\One\otimes\One\\
 &=&(id\otimes id\otimes C_i)(\Delta\otimes id)\Delta+(id\otimes C_i\otimes id)(\Delta\otimes\One)+C_i\otimes\One\otimes\One.
\end{eqnarray*}
On the other hand,
\begin{eqnarray*}
(id\otimes\Delta)\Delta C_i&=&(id\otimes\Delta)((id\otimes C_i)\Delta+C_i\otimes\One)\\
&=&(id\otimes\Delta C_i)\Delta+C_i\otimes\One\otimes\One\\
&=&id\otimes((id\otimes C_i)\Delta+C_i\otimes\One)\Delta+C_i\otimes\One\otimes\One\\
&=&(id\otimes id\otimes C_i)(id\otimes\Delta)\Delta+(id\otimes C_i\otimes id)(\Delta\otimes\One)+C_i\otimes\One\otimes\One
\end{eqnarray*}
which proves the coassociativity by induction on the grading.
Now setting $\epsilon(\One)=\One$ and $\epsilon=0$ elsewhere finishes the proof.
Note that any connected graded bialgebra is a Hopf algebra in a unique way.$\quad\Box$

\subsection{The Hopf algebra of rooted trees, relation to previous work}
\label{normalhopf}
In this section we collect well known results \cite{CK,CK2,Kreimer,Kreimer2} on Hopf algebra methods in momentum space renormalization which will turn out to be applicable to Epstein-Glaser renormalization as well.\\\\
A \emph{rooted tree} is a connected contractible compact graph with a distinguished vertex, the \emph{root}. A \emph{forest} is a disjoint union of rooted trees. Isomorphisms of rooted trees or forests are isomorphisms of graphs preserving the distinguished vertex/vertices. Let $t$ be a rooted tree with root $o.$ The choice of $o$ determines an orientation of the edges of $t:$ we draw the root on top and let the rest of the tree ``hang down.'' Vertices of $t$ having no outgoing edges are called \emph{leaves}, the other vertices (and the root) are called \emph{internal vertices}. The set of forests is graded, for instance by the number of vertices a forest has (the \emph{weight grading}). \\\\  
Let $\h$ be the free commutative algebra generated by rooted trees with the weight grading. 
The commutative product in $\h$ will be
visualized as the disjoint union of trees, such that monomials in $\h$ are scalar multiples of forests.
We demand that the linear operator $B_+$ on $\h,$ defined by
\def\ocf{
\parbox{16mm}{\begin{picture}(16,15)
\put(8,12){$\bullet$}
\put(1,5){$\bullet$}
\put(15,5){$\bullet$}
\put(9,13){\line(1,-1){7}}
\put(9,13){\line(-1,-1){7}}
\put(9,13){\line(1,-2){3.5}}
\put(9,13){\line(0,-1){7}}
\put(9,13){\line(-1,-2){3.5}}
\put(1,2){$t_1$}
\put(7,2){$\ldots$}
\put(15,2){$t_n$}
\end{picture}}}
\begin{eqnarray*}
B_+(\One)&=&\bullet\\
B_+(t_1 \ldots t_n)&=&\ocf
\end{eqnarray*}
is a Hochschild 1-cocycle, which makes $\h$ a Hopf algebra by virtue of \lref{bplusyieldshopf}. It is easy to see that the resulting coproduct can be described as follows
\begin{equation}\label{cuts}
\Delta (t)=\One\otimes t+t\otimes\One+ \sum_{adm. c} P_c(t)\otimes R_c(t)
\end{equation}
where the sum goes over all \emph{admissible cuts} of the tree $t.$ By a cut of $t$ we mean a nonempty set of edges of $t$ that are to be removed. The product of subtrees which ``fall down'' upon removal of those edges is called the \emph{pruned part} and denoted $P_c(t),$ the part which remains connected with the root $R_c(t).$ Now a cut $c(t)$ is admissible, if for each leaf $l$ of $t$ it contains at most one edge on the path from $l$ to the root. For instance,
\begin{eqnarray*}
\Delta\left(\ocnpo\,\right)&=&\ocnpo\otimes\One+\One\otimes\ocnpo+\oc\otimes\oline+\\
&+&\bullet\otimes\ofork+2\bullet\otimes\occ+\oc\bullet\otimes\bullet+2\bullet\bullet\otimes\olongline+\\
&+&\bullet\bullet\otimes\oc+\bullet\bullet\bullet\otimes\oline.
\end{eqnarray*}
$\h$ is obviously not cocommutative.\\\\
Let $V$ be a unital ring with multiplication $m_V.$ Given ring
homomorphisms $\phi, \psi: \h\rightarrow V,$ one can define their
convolution product $\phi\star\psi: \h\rightarrow V,$ $x\mapsto
m_V(\phi\otimes\psi)\Delta x,$ which is a ring homomorphism again. In particular, the antipode $S$ is the
inverse of $id: \h\rightarrow \h$ with respect to this convolution
product. Let $Q$ be the linear endomorphism of $\h\otimes\h$ such that $Q(\One\otimes\One)=-\One\otimes\One$ and $Q=id\otimes P$ otherwise. So (up to the sign) $Q$ is a projection onto $\h\otimes\ker\epsilon\oplus k\One\otimes k\One.$ The shorthand notation $\phi\star_Q\psi:= m_V(\phi\otimes\psi)Q\Delta$ will be useful.\\\\
Now in any Hopf-algebra approach \cite{Kreimer,Kreimer2,CK,CK2} to perturbative quantum field theory, renormalization boils down to twisting the antipode which, (in any graded Hopf algebra) satisfies the recursive equation
\begin{equation*}
S=-m(S\otimes id)Q\Delta=-S\star_Q id,
\end{equation*}
 by a  homomorphism $\Phi: \h\rightarrow V, $ called ``Feynman rules'',
for example into a ring $V$ of
Laurent series (dimensional regularization) or formal power series
(BPHZ), and a ``renormalization scheme'' $R:
V\rightarrow V$
which delivers the counterterm. More explicitly, one considers
\begin{equation}\label{srphi2}
S_{R}^\Phi:= -Rm_V(S_{R}^\Phi\otimes\Phi)Q\Delta=-R(S_{R}^\Phi\star_Q \Phi).
\end{equation}
While $\Phi$ means application of unrenormalized Feynman rules, the renormalized expression is then given by
\begin{equation}\label{srphi}
S_{R}^\Phi\star\Phi.
\end{equation}
For details the reader is referred to \cite{CK}.
In Epstein-Glaser renormalization, essentially the same happens, but
in an easier way because no regularization is required. The
target ring $V$ is most suitably chosen to be something like the tensor algebra of
distributions on $M,$ $\Phi$ will then map a given ``subdivergence situation''
encoded in a rooted tree to the corresponding distribution in $V.$ The
meaning of $\Phi$ is much easier to understand however if we give a
somewhat different presentation of the Hopf algebra and define a
modified convolution product.
\subsection{The cut product and the Bogoliubov recursion}
We enlarge the Hopf algebra $\h$ to $\htwo$ by allowing for two types of vertices: $\bullet$ and $\ast.$ This yields two Hochschild 1-cocycles $B_{+\bullet}$ and 
$B_{+\ast}$ depending on which type the newly adjoined root has. 
\def\ocfd{
\parbox{16mm}{\begin{picture}(16,15)
\put(8,12){$\ast$}
\put(1,5){$\bullet$}
\put(15,5){$\bullet$}
\put(9,13){\line(1,-1){7}}
\put(9,13){\line(-1,-1){7}}
\put(9,13){\line(1,-2){3.5}}
\put(9,13){\line(0,-1){7}}
\put(9,13){\line(-1,-2){3.5}}
\put(1,2){$t_1$}
\put(7,2){$\ldots$}
\put(15,2){$t_n$}
\end{picture}}}
\begin{eqnarray*}
B_{+\bullet}(\One)=\bullet & & B_{+\ast}(\One)=\ast\\
B_{+\bullet}(t_1 \ldots t_n)=\ocf & & B_{+\ast}(t_1\ldots t_n)=\ocfd
\end{eqnarray*}
It is easy to see that the coproduct $\Delta$ which we endow $\htwo$ with using $B_{+\bullet},$ $B_{+\ast}$ 
and \lref{bplusyieldshopf} has the same form (\ref{cuts}) as in $\h.$  Now let $R$ be the algebra endomorphism of $\htwo$ which changes the type of the root to $\ast,$ 
whatever it was before.
\begin{eqnarray*}
R(\bullet)=\ast,\quad R(\ast)=\ast,\quad R\left(\ocf\,\,\,\right)=\ocfd\,\,.
\end{eqnarray*}
Once again we remark that all our algebra endomorphisms are supposed to be unital, so we won't specify their values at $\One$ explicitly. \\\\
Our aim is now to construct a new product $\odot$ called \emph{cut product} of linear endomorphisms of $\htwo.$ The usual convolution product
\begin{equation*}
(\phi,\psi)\mapsto\phi\star\psi = m(\phi\otimes\psi)\Delta
\end{equation*}
in $End_k(\h)$ or $End_k(\htwo)$ has the disadvantage that, applied several times with the projection $P$ onto the augmentation ideal, it gets rid of the structure of trees. For example, for any tree $t$ there is an  $n\in\N$ such that
\begin{equation*}
P^{\star n}(t)=(P\star\ldots\star P) (t)= \mbox{polynomial in }\bullet.
\end{equation*}
Our new product $(\phi\odot\psi)(t)$ is supposed to apply $\phi$ to $P_c(t)$ and $\psi$ to $R_c(t)$ as well, but reassemble the tree afterwards rather than taking the disjoint union of pruned and root parts using $m$. For instance,
\begin{equation*}
(\phi\odot\psi)\left(\oline\right):=\phi\left(\oline\right)\psi(\One)+\phi(\One)\psi\left(\oline\right)+\olinephipsi
\end{equation*}
which should be compared to
\begin{equation*}
(\phi\star\psi)\left(\oline\right)=\phi\left(\oline\right)\psi(\One)+\phi(\One)\psi\left(\oline\right)+\phi(\bullet)\psi(\bullet).
\end{equation*}
This is however only possible for a rather small class of $\phi$ and $\psi$ which do not change the trees too much. For example, $\phi$ is supposed to map trees to trees while $\psi$ is not allowed to kill the vertices where something has been cut. We leave it to the reader to find the most general notion of those maps, because the only ones we need here are $B_+$ and $id,$ $P,$ $R,$ where all this is possible in a rather trivial way.\\\\
Let $\htwobar$ be the Hopf algebra of trees as in $\htwo$ with an additional decoration of the vertices by subsets of $\N.$ There is an obvious forgetful projection $\pi: \htwobar\rightarrow\htwo$ and an inclusion $j: \htwo\rightarrow\htwobar$ decorating all vertices by the empty set. We lift any of the maps $\phi=B_+,id,P,R:\htwo\rightarrow\htwo$ to a map $\tilde{\phi}:\htwobar\rightarrow\htwobar$ by the prescription that newly created vertices are to be decorated by the empty set while the decorations of the old vertices is to be preserved.\\\\
We consider the map $\tilde{\Delta}:\htwobar\rightarrow\htwobar\otimes\htwobar$ which does the same as $\Delta$ in $\htwo$ but decorates each root in $P_c$ and each vertex in $R_c$ that got separated by a cut by the same integer (by the smallest unused integer say), preserving the existing decoration. For example,

\begin{equation*}
\tilde{\Delta}\left(\oc\right)=\oc\otimes\One+\One\otimes\oc+\bullet_{1}\otimes\olone+\bullet_{2}\otimes\otwo+\bullet_1\bullet_2\otimes\bullet_{12}.
\end{equation*}
Here we do not display the empty set and set brackets for simplicity. Note that we do not contend that $\tilde \Delta$ is a coproduct. The decoration has the only purpose to provide ``glueing'' information.\\\\
We define a map $\tilde{m}: \htwobar\otimes\htwobar\rightarrow\htwobar $ which reconstructs the preimage of $\tilde{\Delta}$ by inserting edges between vertices that have been decorated by the same integers and discards the used decoration afterwards. So $\tilde{m}=\tilde{\Delta}^{-1}$ on the image of $\tilde{\Delta}$ and otherwise, if no decorations match, $\tilde{m}$ is the free multiplication $m_{\htwobar}$ of $\htwobar.$ For instance,
\begin{equation*}
\tilde{m}\left(\bullet_1\bullet_2\bullet_3\otimes\olone\bullet_2\bullet_4\right)=\oc\oline\bullet_3\bullet_4
\end{equation*}
$\tilde{m}$ is obviously not an algebra homomorphism.

\begin{dfn} Let $\phi\in \{id,P,R\}$ and $\psi\in \{id,P,R,B_+\}.$ Then the linear endomorphism $\phi\odot\psi$ of $\htwo,$
\begin{equation*}
(\phi\odot\psi)=\pi\tilde{m}(\tilde{\phi}\otimes\tilde{\psi})\tilde{\Delta}j
\end{equation*}
is called the \emph{cut product} of $\phi$ and $\psi.$
\end{dfn}
It is easy to see that if $\phi$ and $\psi$ are algebra endomorphisms, so is $\phi\odot\psi.$\\\\
As a shorthand notation, we will be using
\begin{equation*}
(\phi\odot_Q\psi):=\pi\tilde{m}(\tilde{\phi}\otimes\tilde{\psi})\tilde{Q}\tilde{\Delta}j
\end{equation*}
where $\tilde Q$ is the obvious lift of $Q$ to $(\htwobar)^{\otimes 2}.$
In analogy to the approach presented in the preceding subsection, we recursively define the twisted antipode by $\tilde{S}_R(\One)=\One$ and
\begin{equation}\label{srtilde}
\tilde{S}_R := -\tilde{R}\tilde{m}(\tilde{S}_R\otimes id)\tilde Q\tilde{\Delta}=-\tilde{R}\tilde{m}(\underbrace{-\tilde{R}\tilde{m}(\ldots\otimes id)\tilde{Q}\tilde{\Delta}}_{\tilde S_R}\otimes id)\tilde Q\tilde{\Delta}.
\end{equation}
Let $S_R:=\pi\tilde{S}_R j.$ If one is willing to ignore the fact that $j\pi\neq id,$ one can view $S_R$ as defined by
\begin{equation*}
S_R := -R(S_R\odot_Q id)
\end{equation*}
which might be a helpful motivation when compared to (\ref{srphi2}).
Note that these are recursive definitions indeed since $\tilde Q\tilde\Delta$ reduces the number of edges and $S_R(\One)=\One$ terminates the recursion.
$S_R$ will turn out to be the counterterm map in the Epstein-Glaser framework. Remember that $R$ is an idempotent algebra endomorphism, hence in particular a Rota-Baxter operator. Therefore $S_R$ and $S_R\odot id$ are algebra endomorphism as well by a general inductive argument \cite{Kreimer4}.

\begin{lem} \label{ren} $(S_R\odot id)B_{+\bullet}=(id-R)B_{+\bullet}(S_R\odot id).$
\end{lem}
\emph{Proof. }
We use the Hochschild closedness of $B_{+\bullet}$,
\begin{equation} \label{hoch2}
\Delta B_{+\bullet}=(id\otimes B_{+\bullet})\Delta+ B_{+\bullet}\otimes\One.
\end{equation}
Now we want to lift this equation to $(\htwobar)^{\otimes 2}$ in order to apply it to $(S_R\odot id):$
\begin{equation} \label{hoch3}
\tilde{\Delta}\bbar=C(id\otimes\tilde B_{+\bullet})\tilde{\Delta}+ \tilde B_{+\bullet}\otimes\One
\end{equation}
where $C$ is a map $\htwobar\otimes\htwobar\rightarrow\htwobar\otimes\htwobar$ which decorates vertices affected by a cut by the same integer. This is the only adjustment we have to make when going from (\ref{hoch2}) to (\ref{hoch3}) because $\tilde{\Delta}j$ and $j\Delta$ differ only by decoration. This yields
\begin{eqnarray*}
(S_R\odot id)B_{+\bullet}&=&\pi\tilde{m}(\tilde{S}_R\otimes id)\tilde{\Delta}jB_{+\bullet}=\pi\tilde{m}(\tilde{S}_R\otimes id)\tilde{\Delta}\tilde B_{+\bullet} j \\
&=&\pi\tilde{m}(\tilde{S}_R\otimes id)(C(id\otimes\tilde B_{+\bullet})\tilde{\Delta}+\tilde B_{+\bullet}\otimes\One)j \\
&=&\pi\tilde{m}(\tilde{S}_R \otimes id)C(id\otimes\tilde B_{+\bullet})\tilde{\Delta}j+\pi\tilde{S}_R\tilde B_{+\bullet} j \\
&=&\pi\tilde{m}(\tilde{S}_R \otimes id)C(id\otimes\tilde B_{+\bullet})\tilde{\Delta}j-\pi \tilde{R} \tilde{m}(\tilde{S}_R\otimes id)\tilde Q\tilde{\Delta}\tilde B_{+\bullet} j \\
&=&\pi\tilde{m}(\tilde{S}_R \otimes id)C(id\otimes\tilde B_{+\bullet})\tilde{\Delta}j-\pi \tilde{R} \tilde{m}(\tilde{S}_R\otimes id)C(id\otimes\tilde B_{+\bullet})\tilde{\Delta}j \\
\end{eqnarray*}
\begin{eqnarray*}
&=&(id-R)\pi\tilde{m}(\tilde{S}_R \otimes id)C(id\otimes\tilde B_{+\bullet})\tilde{\Delta}j \\
&=&(id-R)\pi\tilde{m}C(\tilde{S}_R\otimes id)(id\otimes\tilde B_{+\bullet})\tilde{\Delta}j \\
&=&(id-R)\pi\tilde{m}C(id\otimes\tilde B_{+\bullet})(\tilde{S}_R\otimes id)\tilde{\Delta}j \\
&=&(id-R)\b(S_R\odot id),
\end{eqnarray*}
where we have used (\ref{hoch3}), $Q(id\otimes B_{+\bullet})=id\otimes B_{+\bullet},$ $Q(B_{+\bullet}\otimes\One)=0$ which are obvious, and $(\tilde{S}_R\otimes id)C=C(\tilde{S}_R\otimes id)$ and $\tilde{m}C(id\otimes B_{+\bullet})=B_{+\bullet}\tilde{m}$ which follow from the definition of $C.$ This finishes the proof. $\quad\Box$ \\\\
\newpage
\begin{exa}\label{bigdiag}
We illustrate the action of the map
\begin{equation*}
S_R\odot id=-\pi R\tilde{m}(-R\tilde{m}(\ldots\otimes id)\tilde{Q}\tilde{\Delta}\otimes id)\tilde{\Delta}j
\end{equation*}
 on the two trees $\oline$ and $\oc.$

\please\begin{diagram}
\htwo &&  \oline\\
\dTo<{\tilde{\Delta}j} &&  \\
\htwobar\otimes\htwobar && \oline\otimes\One+\One\otimes\oline+\bullet_{1}\otimes\bullet_{1}\\
\dTo<{\tilde Q\Delta\otimes id}&&\\
(\htwobar)^{\otimes 3} && \left(\One\otimes\oline+\bullet_{1}\otimes\bullet_{1}\right)\otimes\One-(\One\otimes\One)\otimes\oline+\One\otimes\bullet_1\otimes\bullet_1\\
\dTo<{\tilde{S_R}\otimes id\otimes id} &&\\
(\htwobar)^{\otimes 3} && \left(\One\otimes\oline-\ast_{1}\otimes\bullet_1\right)\otimes\One-(\One\otimes\One)\otimes\oline+\One\otimes\bullet_1\otimes\bullet_1\\
\dTo<{-R\tilde{m}\otimes id}&&\\
\htwobar\otimes\htwobar && \left(-\octline+\octbline\right)\otimes\One+\One\otimes\oline-\ast_1\otimes\bullet_1\\
\dTo<{\pi\tilde{m}}&&\\
\htwo && -\octline+\octbline+\oline-\ocbline.
\end{diagram}
Note that we do not need to go into higher than the third tensor power of $\htwobar$ because $S_R(\One)=\One$ and hence $S_R(\bullet)=-\ast$ terminate the recursion. Now the second, less trivial example:
\please\begin{diagram}
\htwo &&  \oc\\
\dTo<{\tilde{\Delta}j} &&  \\
\htwobar\otimes\htwobar && \oc\otimes\One+\One\otimes\oc+\bullet_{1}\otimes\olone+\bullet_{2}\otimes\otwo+\bullet_{1}\bullet_{2}\otimes\bullet_{12}\\
\dTo<{\tilde Q\Delta\otimes id}&& \\
(\htwobar)^{\otimes 3} && \left(\One\otimes\oc+\bullet_{1}\otimes\olone+\bullet_{2}\otimes\otwo+\bullet_{1}\bullet_{2}\otimes\bullet_{12}\right)\otimes\One-\One\otimes\One\otimes\oc\\
&& +\One\otimes\bullet_1\otimes\olone+\One\otimes\bullet_2\otimes\otwo+(\One\otimes\bullet_1\bullet_2+\bullet_1\otimes\bullet_2+\bullet_2\otimes\bullet_1)\otimes\bullet_{12}\\
\dTo & &
\end{diagram}
\begin{diagram}
&&\\ \dTo<{\tilde{S_R}\otimes id^{\otimes 2}} &&\\
(\htwobar)^{\otimes 3} && \left(\One\otimes\oc-\ast_{1}\otimes\olone-\ast_{2}\otimes\otwo+\ast_{1}\ast_{2}\otimes\bullet_{12}\right)\otimes\One-\One\otimes\One\otimes\oc\\
 && +\One\otimes\bullet_1\otimes\olone+\One\otimes\bullet_2\otimes\otwo+(\One\otimes\bullet_1\bullet_2-\ast_1\otimes\bullet_2-\ast_2\otimes\bullet_1)\otimes\bullet_{12}\\
\dTo<{-R\tilde{m}\otimes id}&&\\
\htwobar\otimes\htwobar && \left(-\octop+\oclefttop+\ocrighttop-\ocleftrighttop\right)\otimes\One+\One\otimes\oc\\
 && -\ast_1\otimes\olone-\ast_2\otimes\otwo+\ast_1\ast_2\otimes\bullet_{12}\\
\dTo<{\pi\tilde{m}}&&\\
\htwo && -\octop+2\oclefttop-\ocleftrighttop+\oc-2\ocleft+\ocleftright.
\end{diagram}

\end{exa}

\subsection{An alternative presentation of the Hopf algebra}
In this subsection we give a somewhat different presentation $\ul\h$ of $\h$ which will turn out to be more instructive for Epstein-Glaser renormalization. The basic idea is as follows: We consider a tree $t$ of the preceding subsections as a trunk and let two more branches, called ``hair'', grow out of each leaf and one more branch out of each unary vertex of the trunk. This yields a tree $\ul t$ in the presentation $\ul\h.$
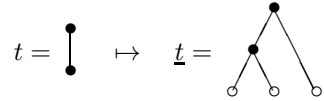
\begin{equation*}
t=\oline\quad\mapsto\quad \ul t= \ocnp
\end{equation*}
\marg
While the trunk will correspond to an abstract nest of subdivergences, the leaves of the hairy tree actually represent (some unordered set of) space-time points to which that particular subdivergence situation applies. For the reader's convenience, we visualize hair by $\circ$ and the trunk vertices by $\bullet.$ This is only to make it easier to distinguish between the bold trees in $\h$ and the hairy trees in $\ul \h,$ so we are not talking about trees with ``two types of vertices'' here. Now in order to underline the power of the Hochschild
1-cocycle and to illustrate \lref{bplusyieldshopf}, we will prescribe the
cocycle and see what the coproduct looks like then. \\\\
Let $\ul\h$ be the free commutative algebra generated by
 rooted trees the leaves of which descend exclusively from binary vertices. In other words each leaf must have one and only one sibling (which is not necessarily a leaf too).
For example, the trees
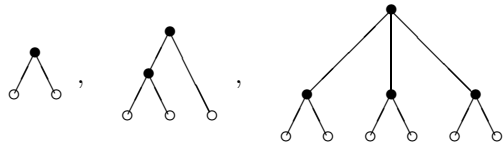
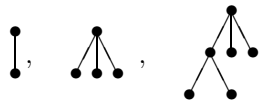
\begin{equation*}
\ocn, \quad\ocnp, \quad\osixn
\end{equation*}
are in $\ul\h$ while
\begin{equation*}
\oline,\quad \othree,\quad\ofoura
\end{equation*}
are not. The tree $\bullet$ consisting only of the root is \emph{not} in $\ul\h$ by convention, so the most ``primitive'' generator is
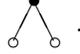
\begin{equation*}
\ocn.
\end{equation*}
Now we demand $\ul B_+$ to act as follows:
\begin{eqnarray*}
\ul B_+(\One)&=&\ocn\\
\ul B_+(\ocn)&=&\ocnp\\
\mbox{in general, for any tree }\ul t,\quad
\ul B_+(\ul t)&=&\oct,\mbox{ so }\ul t\mbox{ is grafted to a leaf of }\ocnr\\
\mbox{and for a forest, }
\ul B_+({\ul t}_1 \ldots {\ul t}_n) &=&\ocful
\end{eqnarray*}
\begin{lem}
There is a unique Hopf algebra structure $(\ul\Delta,\ul\epsilon,\ul S)$ on $\ul\h$ such that $\ul B_+$
is Hochschild closed. $\ul\Delta$ is given on trees $\ul t$ by
\begin{equation*}
\ul\Delta(\ul t)=\One\otimes \ul t+\ul t\otimes\One+\sum_{adm' c}\ul P_c(\ul t)\otimes \ul R_c(\ul t)
\end{equation*}
where the definition of admissible cuts and $\ul P_c,$ $\ul R_c$ is as in the preceding subsections with the following modifications: 
\begin{enumerate}
\item[(i)] cuts containing external edges (hair) are not admissible here
\item[(ii)] if a vertex $v$ of $R_c(\ul t)$ has no more outgoing edges due to cut edges in $c,$ that vertex $v$ is to be replaced by $\ocn$ in $\ul R_c(\ul t).$\\ If a vertex $v$ of $R_c(\ul t)$ is left with only one outgoing edge due to cut edges in $c,$ an additional branch is to be adjoined to $v$ in $\ul R_c(\ul t).$
\end{enumerate}
The map $\beta: \ul\h\rightarrow\h,$ given by removing the hair, i.~e.~ all leaves and adjacent edges, is an isomorphism of Hopf algebras. $\beta^{-1}$ in turn replaces vertices with fertility $0$ or $1$ by binary vertices.
\end{lem}
\emph{Sketch of proof. }First of all we note that whole $\ul\h-k\One$ is the iterated image of $\ul B_+$ and the multiplication. Moreover, $\ul \h$ is graded as an algebra by the number of internal (non-hairy) vertices. The existence and uniqueness of $(\ul \Delta,\ul\epsilon,\ul S)$ is then a consequence of \lref{bplusyieldshopf}.
The remaining statements are easy to check using the map $\beta,$ in particular
\begin{equation*}
\beta\left(\ocn\right)=\bullet, \quad\beta\left(\ocnp\right)=\oline.\quad\quad\Box
\end{equation*}
Therefore $\ul\h$ is nothing but a somewhat different presentation of $\h.$ Using $\beta,$ we can transfer all notions developped in the preceding subsections to $\ul\h$ (which we denote by underlining everything).
Note that in $\mathcal{\ul H}^{\bullet\ast}$ only \emph{internal} vertices can have type $\ast,$ in $\tilde{\mathcal{\ul H}}^{\bullet\ast}$ only \emph{internal} vertices are decorated etc. 
From now on, we work only in the presentation $\ul\h,$ $\mathcal{\ul H}^{\bullet\ast}$, $\tilde{\mathcal{\ul H}}^{\bullet\ast}$.

\subsection{Feynman rules and counterterms. Main result}
On the Hopf algebra level, a tree represents a certain subdivergence situation. Internal vertices of type $\bullet$ mean that the unrenormalized Feynman rules have been applied to the respective subdivergence, while $\ast$ denotes the corresponding counterterm. For example,
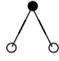
\begin{equation*}
\ocn
\end{equation*}
corresponds to the distribution $^0T_2: f_1\otimes f_2 \mapsto (p_{1,2}T_1\otimes T_1)(f_1\otimes f_2)+(p_{2,1}(T_1\otimes T_1)(f_2\otimes f_1),$ defined on $M^2-\Delta_2.$ Again we do not display the Wick multiindex $r$ for simplicity. The tree
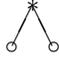
\begin{equation*}
\ocntop
\end{equation*}
represents the counterterm $\left.-W^{\ast}_{12}\right.^0T_2.$ We already know that their sum $(id-W^{\ast}_{12})^0T_2=T_2$ is the well defined Epstein-Glaser time-ordered product on \emph{whole} $M^2.$ In less trivial cases subtrees represent subdivergences, the root represents the overall divergence. For example
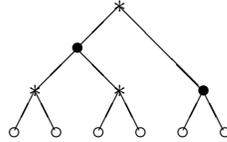
\begin{equation*}
\ocfive
\end{equation*}
yields
\begin{equation*}
W_{123456}^{\ast}p_{1234,56}p_{12,34}W_{12}^{\ast}p_{1,2}W^{\ast}_{34}p_{3,4}p_{5,6}T_1^{\otimes 6}+\mbox{suitable perm. of indices.}
\end{equation*}
Epstein-Glaser renormalization is essentially a binary operation since in each step only products $T(I)T(N-I)$ of \emph{two} operator-valued distributions are considered. Indeed, it is impossible to extend a distribution from $M^n-D_n$ (for $n>2$) onto the thin diagonal in $(M^n-D_n)\cup\Delta_n$ without extending it to the thicker diagonals, e.~g.~ $\{x_i=x_j$ for some $i,j\}$ first. So we will be needing only binary trees here.\\\\
Now let $\ul t$ be a binary tree in $\ul \h.$ All of its internal vertices are of type $\bullet.$ We need a map which changes the types of internal vertices of $\ul t$ in all possible combinations and sums up the resulting trees in order to take care of the Bogoliubov recursion. This is essentially done by $\ul S_R\odot id,$ as we have proved in \lref{ren}. In order to avoid overcounting, we will have to take care of the symmetry factors which show up whenever the coproduct is applied. For instance, in the second part of \eref{bigdiag} we got $2\oclefttop$ because two cuts, one on the ``left'', the other on the ``right hand side'', yield the same result. We will compensate that by eventually 
dividing by symmetry factors. \\\\
Let $\ul{\mathcal{T}}_1=\{\One\}$ and for $n\ge 2$ let $\ul{\mathcal{T}}_n$ be the subset of $\ul\h$ of \emph{binary} trees with $n$ leaves.
Furthermore, let $FV$ be the free commutative algebra generated by the graded vector space
\begin{equation*}
V := \bigoplus_{n=0}^\infty \mathcal{D}'_{op}(M^n-D_n)
\end{equation*}
where $\mathcal{D}'_{op}(M^n-D_n)$ is the space of collections $(T_n^r)$ of operator-valued distributions on $M^n-D_n$ (again $r$ is an $n$-multiindex referring to the Wick powers under consideration). By $D_n$ we continue to denote the thick diagonal in $M^n.$ Thus elements of $FV$ are formal free commutative products of operator-valued distributions on the configuration spaces $M^n-D_n$ carrying a Wick-multiindex. The free commutative product is supposed to model the analogue of the disjoint union of trees. We don't actually need it to state the theorem, but it is instructive to keep it in mind. The reader might wish to review the notation for Epstein-Glaser time-ordered products in subsection 1.2 at this point. 

\begin{thm}[Main result] \label{main}
Let $\Phi: \mathcal{\ul H} ^{\bullet\ast}\rightarrow FV$ be the homomorphism of free commutative algebras such that
\begin{equation*}
\Phi(\One)=T_1\mbox{ where }T_1^k=:\phi^k:
\end{equation*}
and for $n\ge 2,$ $1\le i\le n-1,$ $\ul t_i\in\mathcal{\ul T}_i,$ $\ul t_j\in\mathcal{\ul T}_{n-i}$ and $f_1\ldots f_n\in \mathcal{D}(M)$ such that $\bigcap_{i=1}^n\mbox{supp }f_i =\emptyset,$ $\Phi(B_{+\bullet}(\ul t_i\ul t_j))$ is the collection of distributions defined by
\begin{equation*}
\Phi(B_{+\bullet}(\ul t_i \ul t_j))(f_1\otimes\ldots\otimes f_n)=
\end{equation*}
\begin{equation*}
=\frac{1}{S(\ul t_i,\ul t_j)}\sum_{I\subset N, |I|=i}p_{I,N-I}\Phi(\ul t_i)(\otimes_{k\in I} f_k)\Phi(\ul t_j)(\otimes_{l\in N-I} f_l)+
\end{equation*}
\begin{equation*}
+p_{N-I,I}\Phi(\ul t_j)(\otimes_{l\in N-I} f_l)\Phi(\ul t_i)(\otimes_{k\in I} f_k),
\end{equation*}
\begin{eqnarray*}
\Phi(B_{+\ast}(\ul t_i \ul t_j))&=&W^{\ast}_{1\ldots n}\Phi(B_{+\bullet}(\ul t_i \ul t_j)).
\end{eqnarray*} while $\Phi(\ul t')=0$ on non-binary trees $\ul t'.$ The symmetry factor $S(\ul t_i,\ul t_j):=2$ if the root of $\ul t_i$ has type $\bullet$ and $\ul t_j=\ul R(\ul t_i),$ and $S(\ul t_i,\ul t_j):=1$ otherwise.\\
Using these renormalized Feynman rules $\Phi,$ the $n$-th Epstein-Glaser time-ordered product is (the unique extension onto $M^n$ of)
\begin{equation}\label{hurra}
T_n:=\sum_{\ul t\in\mathcal{\ul T}_n}\Phi(\ul S_R\odot id)(\ul t).
\end{equation}
\end{thm}
Note that in an obvious abuse of notation we consider the counterterms as distributions on $M^n-D_n$ too. Recall that the extension onto $M^n$ is only unique up to the $\omega_\alpha$ as discussed in \cref{uniqueness}. We assume here that for each $n,$ those $\omega_\alpha$ have been chosen once and forever according to some renormalization scheme. \\\\
\emph{Proof.} For $n=1$ and $n=2$ the statement is obviously true (take $\ul t_1=\ul t_2=\One).$ Now for $\ul t\in\mathcal{\ul T}_n$ it is easy to see that $(\Phi \ul R)(\ul t)=(W^{\ast}_{1\ldots n}\Phi)(\ul t)$ (note that $W^\ast$ is idempotent as well) and $\Phi \ul B_{+\bullet}$ is the very sum of causal partitions times lower order time-ordered products that shows up in the equation
\begin{equation}\label{eg2}
T_n = (id-W^{\ast}_{1\ldots n})\sum_{\emptyset\subsetneq I\subsetneq N}p_{I,N-I}T(I)T(N-I)
\end{equation}
which defines the time-ordered product $T_n$ by \cref{uniqueness}. Symbolically, the diagrams
\please\begin{diagram}
\mathcal{\ul H}^{\bullet\ast} & \rTo_\Phi & FV & \quad\quad\quad\quad\quad\quad& \mathcal{\ul H}^{\bullet\ast} & \rTo_\Phi & FV\\
\dTo<{\ul B_{+\bullet}} & & \dTo>{\times\sum p_{I,N-I}\ldots} & & \dTo<{\ul R} & & \dTo>{W^\ast}\\
\mathcal{\ul H}^{\bullet\ast} & \rTo_\Phi & FV & \quad\quad\quad\quad\quad\quad& \mathcal{\ul H}^{\bullet\ast} & \rTo_\Phi & FV
\end{diagram}
commute. This can be seen as follows:
\begin{equation*}
\mathcal{\ul T}_n=\ul B_+(\mathcal{\ul T}_{n-1})\cup \bigcup_{i=2}^{n-2}\ul B_+(\mathcal{\ul T}_i\mathcal{\ul T}_{n-i})=\bigcup_{i=1}^{n-1}\ul B_+(\mathcal{\ul T}_i\mathcal{\ul T}_{n-i})
\end{equation*}
where we are overcounting since $\ul \h$ is commutative. Using the Hochschild closedness of $B_+$ in the form of \lref{ren} and the fact that $\ul S_R\odot id$ is an algebra homomorphism, we get by induction on $n,$ using the symmetry factor $S'(\ul t_i,\ul t_j):=2$ if $\ul t_i=\ul t_j$ and $S'(\ul t_i,\ul t_j):=1$ otherwise:
\begin{eqnarray*}
T_n &=&\sum_{\ul t\in\mathcal{\ul T}_n}\Phi(\ul S_R\odot id)(\ul t)\\
&=&\frac{1}{2}\sum_{i=1}^{n-1}\sum_{\ul t_i\in\mathcal{\ul T}_i}\sum_{\ul t_j\in\mathcal{\ul T}_{n-i}}S'(\ul t_i,\ul t_j)\Phi(\ul S_R\odot id)\ul B_+(\ul t_i\ul t_j)\\
&=&\frac{1}{2}\sum_{i=1}^{n-1}\sum_{\ul t_i\in\mathcal{\ul T}_i}\sum_{\ul t_j\in\mathcal{\ul T}_{n-i}}S'(\ul t_i,\ul t_j)\Phi(id-\ul R)\ul B_{+\bullet}(\ul S_R\odot id)(\ul t_i \ul t_j)\\
&=&\frac{1}{2}(id-W^{\ast}_{1\ldots n})\sum_{i=1}^{n-1}\Phi \ul B_{+\bullet}\left(\sum_{\ul t_i\in\mathcal{\ul T}_i}(\ul S_R\odot id)(\ul t_i)\sum_{\ul t_j\in\mathcal{\ul T}_{n-i}}(\ul S_R\odot id)(\ul t_j)+C\right)\\
\end{eqnarray*}
\begin{eqnarray*}
&=&\frac{1}{2}(id-W^{\ast}_{1\ldots n})\sum_{i=1}^{n-1}\sum_{I\subset N, |I|=i}p_{I,N-I}T(I)T(N-I)+p_{N-I,I}T(N-I)T(I)\\
&=&(id-W^{\ast}_{1\ldots n})\sum_{\emptyset\subsetneq I\subsetneq N}p_{I,N-I}T(I)T(N-I)
\end{eqnarray*}
where $C$ is eventually $C=\sum_{\ul t} (\ul S_R\odot id)(\ul t)(\ul S_R\odot id)(\ul t)$ (for each
$\ul t$ such that $\ul t_i=\ul t_j=:\ul t$ has occurred in the sum above, thus in particular for all $\ul t\in \mathcal{\ul T}_{n/2}$ if $n$ is even) which cancels the symmetry factor $S(\ul t_i,\ul t_j)$ in the statement of the theorem.
This finishes the proof. $\quad\Box$\\\\
While the preceding theorem just defines $\Phi$ inductively by pushing it forward along $\ul B_+,$ which is a perfectly natural way of doing so, one might also work out a non-recursive formula for $\Phi$ as follows: Draw the tree, scan it from the top to the bottom and wherever you see an $\ast,$ apply $W^\ast.$ Then symmetrize in all possible ways.\\\\
Since $\ul\h$ is nothing but a different presentation of $\h,$ one could also have stated the theorem in terms of 
trees of $\h$ from the very beginning, which would have required a grading on $\h$ that is isomorphic to the grading of $\ul \h$ by the number of external (hairy) vertices.\\\\
We encourage the reader to check that one could obtain the same result in \emph{complete} analogy to 
momentum space renormalization (BPHZ, dimensional regularisation, etc.) \cite{CK,CK2,Kreimer,Kreimer2} as 
reviewed in subsection \ref{normalhopf} by the following approach: Define the (unrenormalized) Feynman rules $\Phi: \ul H\hookrightarrow \ul{H}^{\bullet\ast}\rightarrow FV$ 
as in \tref{main}, but let now $R: FV\rightarrow FV$ be 
the idempotent algebra endomorphism $T\mapsto W^\ast T.$ 
Note that $R$ is a Rota-Baxter operator. Then replace the cut product $\odot$ by the usual convolution product $\star$ again, and an adaptation of \tref{main} yields
\marg
\begin{equation*}
T_n = \sum_{\underline{t}\in \mathcal{\ul T}_n}(\ul S_{R}^\Phi\star \Phi)(\ul t)
\end{equation*}
which should be compared to (\ref{srphi}).
The reason why we preferred the method of letting $R$ act in the Hopf algebra 
$\htwo$ and using $\odot$ is that like this we achieved a complete decoupling of the 
combinatorics (which happen in $\htwo$) and the analysis (which happens in $V$), 
making it easier to see how the essential work is being done on the Hopf algebra side while the renormalized 
Feynman rules $\Phi: \mathcal{\ul H}^{\bullet\ast}\rightarrow FV$ 
is a rather trivial map translating abstract subdivergence situations into the appropriate operator valued distributions.
\section{Conclusions and outlook}
We have seen how Hopf algebras of rooted trees take care of the combinatorics of Epstein-Glaser renormalization. It is the twisted antipode $S_R$ which provides a complete set of counterterms and formally solves the Bogoliubov recursion thanks to the Hochschild closedness of $B_+.$ The statement of \lref{ren} also amounts to the fact that the counterterms are local. Indeed, once the subdivergences are taken care of, it suffices to subtract the superficial divergence, i.~e.~ to extend a distribution onto the thin diagonal. Although we do not claim that the statement of \tref{main} makes actual calculations easier, it closes the gap between the Epstein-Glaser approach and the Hopf algebra picture in momentum space.  \\\\
Starting from \tref{main}, one rather easily derives Feynman rules $\Phi$ for the \emph{vaccum expectation values} of time-ordered products. One can also try to construct a coproduct on the vacuum expectation values of time-ordered products.\\\\ 
Finally, we would like to mention another issue which seems to be intimately related to the above approach to coordinate space renormalization: constructing an analogy between extension of distributions from $M^n-D_n$ to $M^n$ and \emph{compactification} of the configuration space $M^n-D_n$ of $n$ points in $M.$ Indeed, we can already see how this leads to rooted trees if we look at the Fulton-MacPherson compactification of configuration spaces \cite{FM,AS,GJ} defined as follows: \\\\
Let $M$ be a  smooth manifold. There is an obvious inclusion of the configuration space into a product of blowups,
\begin{equation}\label{fmincl}
M^n-D_n\hookrightarrow M^n\times\prod_{I\subset N, |I|\ge 2}Bl(M^{|I|},\Delta_{|I|})
\end{equation}
 where $Bl(M^i,\Delta_i)$ is the (differential-geometric) blowup of $M^i$ along $\Delta_i$ of $M^i,$ i.~e.~ $M^i$ where the thin diagonal $\Delta_i$ is replaced by the sphere bundle in the normal bundle over $\Delta_i.$ For the details, the reader is referred to \cite{AS}. The \emph{Fulton-MacPherson compactification} $M[n]$ of $M^n-D_n$ is then the closure of $M^n-D_n$ upon this inclusion. Obviously $M[n]$ has only a chance to be compact if $M$ is compact. Now a closer look at what happens in the right hand side of (\ref{fmincl}) when a sequence approaches the thin diagonal in $M^n$ leads to a nice description of $M[n]$ in terms of nested \emph{screens} \cite{FM,AS}. In particular, it can be shown that there is a stratification of the manifold with corners $M[n],$
\begin{equation*}
M[n]=\bigcup_{S\in\mathcal{S}} M(S)
\end{equation*}
where $\mathcal{S}$ is the set of all nests of subsets of $N=\{1,\ldots,n\}$ with at least 2 elements. Now nested sets are perfectly described by the forests in $\h.$ Moreover, if we restrict our attention to $M=\R^k$ and replace $M^n-D_n$ by the moduli space $\dot F_k(n):= (M^n-D_n)/G(k)$ where $G(k)$ is the subgroup (acting diagonally) of affine transformations of $\R^k$ generated by translations by elements of $\R^k$ and dilatations by elements of $\R_+,$ there is an \emph{operad} structure behind the Fulton-MacPherson compactifications $F_k(n)$ of the moduli spaces $\dot F_k(n)$ \cite{GJ,MSS}. The compactifications $M[n]$ of the configuration spaces still furnish a right module over the operad $F_k.$ Operads arise in a natural way when rooted trees are grafted to each other:
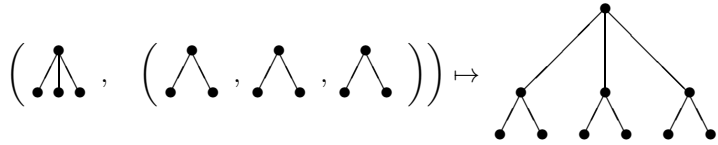
\begin{equation*}
\left(\othree,\quad \left(\oc,\oc,\oc\right)\right)\mapsto \osixend
\end{equation*}
\marg
It seems tempting to explore possible relations between the operad $\mu_{FM}$ of Fulton-Mac\-Pherson compactified moduli spaces $F_k(n),$ the operad $\mu_{EG}$ which arises when the trees in $\h$ we used for Epstein-Glaser renormalization are grafted to each other, and finally the operad of Feynman graph insertions $\mu_{FG}$ \cite{MSS,Kreimer3}. The operad $\mu_{FG}$ is closely related to the pre-Lie structure of Feynman graphs which is dual in a certain sense to the coproduct in $\h.$ This might establish a true analogy between the Fulton-MacPherson compactification $M[n]$ of $M^n-D_n$ and the renormalization of time-ordered products in the sense of Epstein-Glaser.

\section*{Acknowledgements}
It is a pleasure to thank Henri Epstein and Ivan Todorov for valuable discussion and helpful comments during a series of talks given on the subject. The first named author would also like to thank the IHES for generous hospitality and the German Academic Exchange Service (DAAD) for financial support.

\end{document}